\documentclass{article}
\usepackage{spconf}
\usepackage[francais,english]{babel}
\usepackage{amsfonts}
\usepackage{amstext}
\usepackage{amssymb}
\usepackage{graphics}
\usepackage{spconf}
\usepackage{epsfig,amsmath}
\name{\textit{Ziad NAJA$^{(1)}$, Florence ALBERGE$^{(1)}$, Pierre
DUHAMEL$^{(2)}$}
\thanks{Thanks to Newcom++ WPR4 for funding.}}
\address{\begin{tabular}{c}
Laboratoire des signaux et syst\`{e}mes (L2S)\\Univ Paris-Sud$^{(1)}$, CNRS$^{(2)}$\\ Supelec, 3 rue Joliot-Curie 91192 Gif-sur-Yvette cedex (France)\\
E-mails: \textit{\{naja, alberge, pierre.duhamel\}@lss.supelec.fr}
\end{tabular}}
\title{Geometrical interpretation and improvements of the Blahut-Arimoto's algorithm}
\begin{document}
\ninept \maketitle \normalfont
\begin{abstract}
The paper first recalls the Blahut Arimoto algorithm for computing
the capacity of arbitrary discrete memoryless channels, as an
example of an iterative algorithm working with probability density
estimates. Then, a geometrical interpretation of this algorithm
based on projections onto linear and exponential families of
probabilities is provided. Finally, this understanding allows also
to propose to write the Blahut-Arimoto algorithm, as a true proximal
point algorithm. it is shown  that the corresponding version has an
improved convergence rate, compared to the initial algorithm, as
well as in  comparison with other improved versions.
\end{abstract}
\section{Introduction}
In 1972, R. Blahut and S. Arimoto \textit{\cite{Arimoto,Blahut}}
received the Information Theory Paper Award for their Transactions
Papers on how to compute numerically the capacity of memoryless
channels with \textbf{\textit{finite}} input and output alphabets.

The Blahut-Arimoto algorithm was recently extended to channels with
memory and finite input alphabets and state spaces
\textit{\cite{Dupuis04}}.

Recently, an algorithm was proposed for computing the capacity of
memoryless channels with \textbf{\textit{continuous}} input and/or
output alphabets where the Blahut-Arimoto algorithm is not directly
applied \textit{\cite{Dauwels}}.

In \textit{\cite{Csiszar_proj}}, information geometric
interpretation of the Blahut-Arimoto algorithm in terms of
alternating information projection was provided. Based on this last
approach, Matz \textit{\cite{Matz04}} proposed a modified
Blahut-Arimoto  algorithm that converges significantly faster than
the standard one.\\The algorithm proposed by Matz is based on an
approximation of a proximal point algorithm. Instead, we propose a
true proximal point reformulation that permits to accelerate the
convergence speed compared to the classical Blahut-Arimoto algorithm
and also to the approach in \textit{\cite{Matz04}}.

Our contributions regarding capacity computation for discrete
memoryless channels (DMCs) in this paper are:

\begin{itemize}
\item Geometrical interpretation of Blahut-Arimoto algorithm in terms of projection onto linear and exponential families of probability.
\item \textbf{\emph{True}} proximal point interpretation.
\item Improvement of the convergence rate based on the proximal point formulation.
\end{itemize}
\section{Tools}
\subsection{Kullback-Leibler divergence and Mutual Information}
The Kullback-Leibler divergence (KLD) \textit{\cite{Cover,Gallager}}
is defined for two probability distributions $ \textit{p}=\{p(x),x
\in \textit{\textbf{X}}\}$ and $\textit{q}=\{q(x),x \in
\textit{\textbf{X}}\}$ of a discrete random variable \textbf{X}
taking their values \textbf{x} in a discrete set \textit{\textbf{X}}
by:
\begin{displaymath}
\textit{D}(\textit{p}||\textit{q})=\sum_{\textbf{x}\in\textit{\textbf{X}}}{p(\textbf{x})\log\frac{p(\textbf{x})}{q(\textbf{x})}}
\end{displaymath}
The KLD(also called relative entropy) has some of the properties of
a metric: $D(p||q)$ is always non-negative, and is zero if and only
if $p=q$. However, it is not a true distance between distributions
since it is not symmetric ($D(\textit{p}||\textit{q})\neq
D(\textit{q}||\textit{p})$) and does not satisfy the triangle
inequality in general. Nonetheless, it is often useful to think of
relative entropy as a "distance" between
distributions.\\
\begin{figure}[!h]
\centerline{\epsfxsize=3.5cm\epsfysize=0.6cm\epsfbox{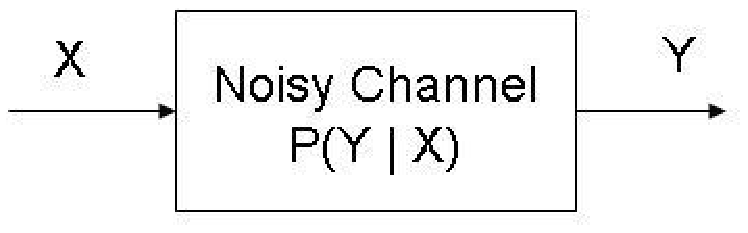}}\vspace*{-0.4cm}\caption{\footnotesize
\label{channel} \tiny {Channel model}}
\end{figure}
The channel capacity is given by:
\[C=\max_{p(x)}{\textit{I(X,Y)}}\]
Where the mutual information of the two discrete random variables
\textbf{X} and \textbf{Y} is given by :
\begin{displaymath}
\textit{I(X,Y)}=\mathbb{E}_p\{D(\textit{p(y$|$x)}||\textit{p(y)})\}
\end{displaymath}
\subsection{Linear and exponential families of probability}
A linear family of probability is defined as
\textit{\cite{Csiszar_proj}} :

$\forall f_1,f_2,\ldots,f_K \in \textbf{X}$ and $\forall
\alpha_1,\alpha_2,\ldots,\alpha_K $
\begin{center}
$\mathcal{L}=\{p:\mathbb{E}_p{(f_i(x))}=\alpha_i,1\leq i\leq K\}$
\end{center}
The expected value $\mathbb{E}_p{(f_i(x))}$ of the random variable x
with respect to the distribution $p(x)$ is restricted to $\alpha_i$.
A linear family of probability is characterized by
$\{f_i(x)\}_{1\leq i\leq K}$ and $\{\alpha_i\}_{1\leq i \leq
K}$.\\The vector $\alpha=[\alpha_1,\ldots,\alpha_k]$ serves as a
coordinate system in the manifold of the linear family. These4
coordinates are called "mixture coordinates".

An exponential family \textit{\cite{Csiszar_proj}} of discrete
probability distributions $p(x)$ on an alphabet \textit{X} is the
set
\begin{center}
$\mathcal{E}=\{p:p(x)=\frac{Q(x)\exp{\sum_{i=1}^{K}(\theta_i
f_i(x))}}{\sum_x(Q(x)\exp{\sum_{i=1}^{K}(\theta_i f_i(x))})}\}$
\end{center}
The exponential family $\mathcal{E}$ is completely defined by
$f_i(x)$ and $Q(x)$ and parameterized by $\theta_i$.\\The
distribution $Q(x)$ is itself an element of the exponential family.
Any element of $\mathcal{E}$ could play the role of Q(x), but if it
is necessary to emphasize the dependence of $\mathcal{E}$ on $Q(x)$,
we will write $\mathcal{E}_Q$.
\section{Blahut-Arimoto-Type Algorithm }
\subsection{The original Blahut-Arimoto algorithm}
Let consider the case of a discrete memoryless channel with input
symbol X taking its values in the set $\{x_0,\ldots,x_M\}$ and
output symbol Y taking its values in the set $\{y_0,\ldots,y_N\}$.
This channel is defined by its transition probabilities channel
matrix Q as $[Q]_{ij}$ = $Q_{i|j}=Pr(Y=y_i|X=x_j)$. We also define
$p_j=Pr(X=x_j)$ and $q_i=Pr(Y=y_i)$.

The mutual information is given by:
\[\textit{I(X,Y)}=\textit{I(p,Q)}=\sum_{j=0}^{M}\sum_{i=0}^{N}{p_j
Q_{i|j} \log\frac{Q_{i|j}}{q_i}}=\sum_{j=0}^{M}{p_j D(Q_j||q)}\]
And the channel capacity by:
\[C=\max_{p}{I(p,Q)}\]

By solving this maximization problem and taking into consideration
the normalization condition: $\sum_x{p(x)}=1$, we find:
\begin{center}
$p_j=\frac{p_j \exp(D(P(Y|X=x_j)||P(Y)))}{\sum_j{p_j[
\exp(D(P(Y|X=x_j)||P(Y)))]}}$
\end{center}
Hence, the Classical Blahut-Arimoto algorithm \cite{Arimoto,Blahut}
is an iterative procedure:
\begin{equation}
\label{BA} p^{(k+1)}{(x)}=\frac{p^{(k)}{(x)}
\exp(D_x^k)}{\sum_{x}^M{p^{(k)}{(x)} \exp(D_x^k)}}
\end{equation}
with $D_x^k=D(p(Y=y|X=x)||p(Y=y^{(k)}))$.
\subsection{Geometrical Interpretation of Blahut-Arimoto Algorithm}
The Blahut-Arimoto algorithm in (\ref{BA}) can be recalculated as a
minimization problem:
\begin{displaymath}
\left\{ \begin{array}{lll}  \min_p & D(p(x)||p^{(k)}(x))\\
    s.c & I^{(k)}{(p(x))}=\alpha\\
    s.c & \sum_x{p(x)}=1
    \end{array} \right.
\end{displaymath}
where $I^{(k)}{(p(x))}=\mathbb{E}_p\{D(p(y|x)||p^k{(y)})\}$ is the
current capacity estimate at the iteration $k$ and $\alpha$ is
related to the Lagrangian multiplier of this minimization
problem.\\The Lagrangian corresponding to this minimization problem
can be written as follow:
\begin{center}
$\mathfrak{L}=D(p(x)||p^{(k)}(x))-\lambda_1{(I^{(k)}{(p(x))}-\alpha)}-\lambda_2{(\sum_x{p(x)}-1)}$
\end{center}
$\frac{\partial\mathfrak{L}}{\partial p(x)}=0$ $\Rightarrow
\log(p(x))+1-\log(p^{(k)}(x))-\lambda_1{D_x^k}-\lambda_2=0$ and
$p(x)=p^{(k)}{(x)}\exp(\lambda_2-1)\exp(\lambda_1{D_x^k})$\\Taking
into consideration the normalization constraint, we can easily
obtain that $\exp(\lambda_2-1)=\frac{1}{\sum_{x}{p^{(k)}{(x)}
\exp(\lambda_1 D_x^k)}}$ and $p^{(k+1)}{(x)}=\frac{p^{(k)}{(x)}
\exp(\lambda_1 D_x^k)}{\sum_{x}{p^{(k)}{(x)} \exp(\lambda_1
D_x^k)}}$\\
In the following, we will see that this parameter $\lambda_1$ is a
step size parameter which, for convenient values, can accelerate the
convergence speed of the classical Blahut-Arimoto algorithm in which
$\lambda_1=1$.\\
So the Blahut-Arimoto Algorithm can be interpreted as the projection
of $p^{(k)}{(x)}$ onto a linear family of probability $\mathcal{L}$
at the point $p^{(k+1)}{(x)}$ where $\mathcal{L}$ is defined by
$f_1(x)=D_x^k=D(p(y/x)||p^{(k)}{(y)})$ and $\alpha_1^k$ such as
$\mathbb{E}_p(D_x^k)=\alpha_1^k$.

By choosing increasing $\alpha_1^k$, we would ensure that the mutual
information increases from one iteration to the other
($I^{(k+1)}{(p(x))}\geq I^{(k)}{(p(x))}$). However, this quantity is
only implicitly defined in the algorithm and an appropriate choice
is not available.In the following, we show that this problem will be
solved based on a proximal point interpretation that ensures that
the mutual information increases during iterations.

%
%
Note that this linear family of probability is changing from one
iteration to the other.\\On the other hand, the Blahut-Arimoto
algorithm can be interpreted as the projection of a probability
density function (pdf) onto an exponential family of probability
$\mathcal{E}$ defined by $Q(x)=p^{(k)}{(x)}$, $f_1^{(k)}{(x)}=D_x^k$
and parametrized with $\theta_1^{(k)}$ at the point
$p^{(k+1)}{(x)}$.

To do this, we should solve this problem:
\begin{displaymath}
\left\{ \begin{array}{l}  \displaystyle{\min_\theta} D(R(x)||p(x,\theta))\\
    p(x,\theta)=\frac{Q(x)exp{(\theta f_1(x)}}{\sum_x Q(x)exp{(\theta
    f_1(x))}}
    \end{array} \right.
\end{displaymath}
 where $R(x)$ is a certain pdf. We try now to find some interesting
characteristics of $R(x)$. To do this, let solve the minimization
problem given above. $\sum_x{\frac{\partial(R(x)\log p(x))}{\partial
\theta}=0}$ with $\log p(x,\theta)=\log Q(x)+\theta
f_1{(x)}-\log(\sum_x{Q(x)\exp(\theta f_1{(x)})})$\\So
 $\sum_x{R(x)f_1{(x)}-\frac{\sum_x{R(x)}\sum_x{Q(x)f_1{(x)}\exp(\theta
f_1{(x)})}}{\sum_x{Q(x)\exp(\theta f_1{(x)})}}}=0$\\Hence
$\sum_x{R(x)f_1{(x)}-\frac{\sum_x{Q(x)f_1{(x)}\exp(\theta
f_1{(x)})}}{\sum_x{Q(x)\exp(\theta f_1{(x)})}}}\sum_x{R(x)}=0$
leading to $\sum_x{(R(x)-p(x,\theta))f_1{(x)}}=0$ having that
$\sum_x{R(x)}=1$ and $p(x,\theta)=\frac{Q(x)\exp(\theta
f_1{(x)})}{\sum_x{Q(x)\exp(\theta f_1{(x)})}}$.\\We obtain
\begin{center}
$\sum_x(R(x)-p^{(k+1)}{(x)})D_x^k=0$
\end{center}
Which can be reformulated as
\begin{center}
$I(R,Q)=\mathbb{E}_R(D_x^k)=\mathbb{E}_p^{(k+1)}(D_x^k)=I(p^{(k+1)}{(x)},Q)\geq
I(p^{(k)}{(x)})$
\end{center}
Hence the Blahut-Arimoto algorithm can be interpreted as the
projection of pdfs $R(x)$ with higher mutual information than
$I(p^{(k)}{(x)})$ onto an exponential family $\mathcal{E}$ defined
by $Q(x)=p^{(k)}{(x)}$, $f_1^{(k)}{(x)}=D_x^k$ and parameterized by
$\theta_1^{(k)}=1/\lambda_k$ at the point $p^{(k+1)}{(x)}$. Note
that this exponential family is also changing from iteration to
another since Q(x) and $f_1^{(k)}{(x)}$ depends on the iteration.
Here again, an appropriate choice of the parameter for increasing
convergence rate is difficult, because of the implicit definition of
the family. Thus, a proximal point interpretation maximizing
explicitly the mutual information is considered with a given penalty
term.
\subsection{Proximal point interpretation of B.A. and amelioration
in terms of convergence speed}   Following the results above, and
based on a proximal point interpretation, we can solve the problem
stated by the implicit definition of the families. In fact, we
propose a clear equivalence with a true proximal point
interpretation, in which all constants are explicitly defined, thus
allowing to propose convergence rate improvement. It is easily shown
that the Blahut-Arimoto algorithm is equivalent to
\begin{equation}
\label{proxi1}p^{(k+1)}{(x)}=\arg\max_p\{I^{(k)}{(p(x))}-D(p(x)||p^{(k)}{(x)})\}
\end{equation}
In fact, by deriving this expression over $p(x)$ and set it equal to
zero, we find exactly the iterative expression of the Blahut-Arimoto
algorithm.

But till now we cannot say that the Blahut-Arimoto algorithm can be
interpreted as a proximal point method since the cost function
$I^{(k)}{(p(x))}$ depends on the iterations, just like the families
were depending on the iterations. In fact, a true proximal point
algorithm can be written for a maximization problem
\textit{\cite{Vige}} as follow :
\begin{equation}
\label{proxi2}\theta^{(k+1)}=\arg\max_{\theta}\{\xi(\theta)-\beta_k
\|\theta-\theta^{(k)}\|^2\}
\end{equation}
in which $\xi(\theta)$, the cost function to be maximized, is
independent from the iterations, $\|\theta-\theta^{(k)}\|^2$ is a
penalty term which ensures that the update $\theta^{(k+1)}$ remains
in the vicinity of $\theta^{(k)}$ and $\beta_k$ is a sequence of
positive parameters. In \cite{Rockafellar}, Rockafellar showed that
superlinear convergence of this method is obtained when the sequence
${\beta_k}$ converges towards zero.\\The definition of the proximal
point algorithm in (\ref{proxi2}) can be generalized to a wide range
of penalty terms leading to this general formulation:
\begin{center}
$\theta^{(k+1)}=\displaystyle{\arg\max_\theta}\{\xi(\theta)-\beta_k
f(\theta,\theta^{(k)})\}$
\end{center}
where $f(\theta,\theta^{(k)})$ is always non negative and
$f(\theta^{(k)},\theta^{(k)})=0$.\\The mutual information $I(p(x))$
can be expressed as:
\begin{equation}
\label{proxi3}I(p(x))=I^{(k)}{(p(x))}-D(q(y)||q^{(k)}{(y)})
\end{equation}
\vspace{-0.3cm} Introducing (\ref{proxi3}) in (\ref{proxi1}) leads
to
\begin{center}
$p^{(k+1)}{(x)}=\arg\max_p\{I(p(x))-(
D(p(x)||p^{(k)}{(x)})-D(q(y)||q^{(k)}{(y)}))\}$
\end{center}
 This new formulation establishes a clear link with the definition
of the capacity based on the mutual information. However, for a true
proximal pint formulation, we need to show that:
\begin{center}
$D(p(x)||p^{(k)}{(x)})-D(q(y)||q^{(k)}{(y)})\geq 0$
\end{center}
with equality iff $p(x)=p^{(k)}{(x)}$ and $q(y)=q^{(k)}{(y)}$ in
order to prove that the Blahut-Arimoto is a proximal point
algorithm.\\The penalty term
$D(p(x)||p^{(k)}{(x)})-D(q(y)||q^{(k)}{(y)})$ can be rewritten as
$\mathbb{E}_{p(x,y)}{[\log\frac{p(x){\sum_{\tilde{x}}{p(y|\tilde{x})p^{(k)}{(\tilde{x})}}}}{p^{(k)}{(x)}{\sum_{\tilde{x}}{p(y|\tilde{x})p(\tilde{x})}}}]}$.\\
We can also write according to Jensen's inequality
\textit{\cite{Cover}} :
\begin{eqnarray}
\label{penaltyterm}
\mathbb{E}_{(p(x,y)}{[-\log\frac{p^{(k)}{(x)}{\sum_{\tilde{x}}{p(y|\tilde{x})p(\tilde{x})}}}{p(x){\sum_{\tilde{x}}{p(y|\tilde{x})p^{(k)}{(\tilde{x})}}}}]}\\
\geq -\log(\sum_y\sum_x{p(x,y)})=0
\end{eqnarray}
\\ This proves that the Blahut-Arimoto algorithm can be interpreted as a true proximal point
method where the cost function is the true mutual information and
the penalty term reads
\begin{center}
$D(p(x)||p^{(k)}{(x)})-D(q(y)||q^{(k)}{(y)})$
\end{center}
The corresponding proximal point algorithm reads:
\begin{equation}
\label{our}
\begin{array}{lcl}
p^{(k+1)}{(x)} & = & \arg\max_{p(x)}\left\{I(p(x))-\lambda_k(D(p(x)||p^{(k)}{(x)}))\right.\\
& &\left. -D(q(y)||q^{(k)}{(y)})\}\right\}
\end{array}
\end{equation}
where $\lambda_k$ is the step size introduced in order to accelerate
the convergence rate of the classical Blahut-Arimoto algorithm.\\
By deriving this function
\begin{center}
$I(p(x))-\lambda_k{(D(p(x)||p^{(k)}{(x)})-D(q(y)||q^{(k)}{(y)}))}$
\end{center}
 and set it equal to zero we find:
\begin{displaymath}
\begin{array}{lcl}
p^{(k+1)}{(x)} & = & p^{(k)}{(x)}\exp\left\{\sum_y{p(y|x)\log\frac{q(y)}{q^{(k)}{(y)}}}-\frac{1}{\lambda_k}\right.\\
& &\left.
+\frac{1}{\lambda_k}\sum_y{p(y|x)\log\frac{p(y|x)}{q(y)}}\right\}
\end{array}
\end{displaymath}
Here, it is important to note that we can obtain the classical case
by simply replacing $\lambda_k$ by $1$. \\Moreover, we can also
obtain the approach proposed by Matz \cite{Matz04} by intuitively
replacing the probability distribution $q(y)$ in the right hand of
the equation by the same distribution calculated at the previous
iteration ($q^{(k)}{(y)}$). Namely:
\begin{displaymath}
\begin{array}{lcl}
p^{(k+1)}{(x)} & = & p^{(k)}{(x)}\exp\left\{\sum_y{p(y|x)\log\frac{q^{(k)}{(y)}}{q^{(k)}{(y)}}}-\frac{1}{\lambda_k}\right.\\
& &\left.
+\frac{1}{\lambda_k}\sum_y{p(y|x)\log\frac{p(y|x)}{q^{(k)}{(y)}}}\right\}
\end{array}
\end{displaymath}
After normalization, we get $p^{(k+1)}{(x)}=p^{(k)}{(x)}\exp(D_x^k/\lambda_k)$ which is the expression of Matz's approach. This is globally similar to the One-Step-Late algorithm suggested by Green\textit{\cite{Green}} \\
\\We conclude that Matz's approach is based on an approximation of the proximal point
method, but what is lost in comparison with the true proximal point
method is the guarantee that the method converges, since convergence
conditions must be reviewed again.
\\We can write according to (\ref{our}):
\begin{center}
$I(p^{(k+1)}{(x)})-\lambda_k{(D(p^{(k+1)}{(x)}||p^{(k)}{(x)})-D(q^{(k+1)}{(y)}||q^{(k)}{(y)}))}\geq
I(p^{(k)}{(x)})-\lambda_k{(D(p^{(k)}{(x)}||p^{(k)}{(x)})-D(q^{(k)}{(y)}||q^{(k)}{(y)}))}
$
\end{center}
Hence
\begin{center}
$I(p^{(k+1)}{(x)})\geq
I(p^{(k)}{(x)})+\lambda_k{(D(p^{(k+1)}{(x)}||p^{(k)}{(x)})-D(q^{(k+1)}{(y)}||q^{(k)}{(y)}))}$
\end{center}
To ensure the increasing of the mutual information during
iterations, we must have:
\begin{center}
$I(p^{(k+1)}{(x)}) \geq I(p^{(k)}{(x)})$
\end{center}
So that
$\lambda_k{(D(p^{(k+1)}{(x)}||p^{(k)}{(x)})-D(q^{(k+1)}{(y)}||q^{(k)}{(y)}))}\geq
0$ which is true, from (\ref{penaltyterm}) for every $\lambda_k\geq
0$ which is not true in the approach proposed by Matz. In our
method, we choose $\lambda_k$ such that:
\begin{center}
$\max_{\lambda_k}{\lambda_k{(D(p^{(k+1)}{(x)}||p^{(k)}{(x)})-D(q^{(k+1)}{(y)}||q^{(k)}{(y)}))}}$
\end{center}
in which $p^{(k+1)}{(x)}$ and $q^{(k+1)}{(y)}$ depend on
$\lambda_k$.\\This ensures that the difference between
$I(p^{(k+1)}{(x)})$ and $I(p^{(k)}{(x)})$ is as maximum as possible
from one iteration to the other one. Note that this maximization
problem is solved by the conjuguate gradient method which gives the
most convenient value of the step size $\lambda_k$ comparing to the
approach proposed by Matz.\\Note that, in terms of algorithmic
complexity, the updated value of $\lambda_k$ in each iteration requires:\\
(N+M+1) divisions and (N+M) multiplications in Matz's
approach.\\(2N+M+1) divisions, (2N+M+2) multiplications and 2
additions in our case based on the proximal point method.\\Hence,
our method requires less than twice operations per iteration
compared to the approach proposed by Matz, however, it converges
faster (as we can see in the simulation results showed below, the
iteration number is divided by two in the worst case). A compromise
must be established depending on our interests.
\section{Simulation results}
\begin{figure}[!h]
\centerline{\epsfxsize=7cm\epsfysize=4cm\epsfbox{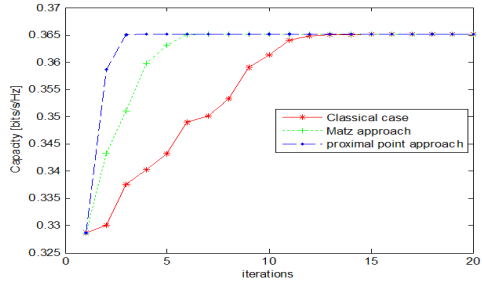}}\vspace*{-0.4cm}\caption{\footnotesize
\label{result1} Comparision between the 3 approaches in the case of
a DBSC channel}
\end{figure}

First, we test the 3 versions of the Blahut-Arimoto iterative
algorithm on a Discrete Binary Symmetric Channel (DBSC) defined by
the transition matrix :
\begin{center}
\[Q= \left \{
\begin{array}{ccc}
   0.7 & 0.2 & 0.1 \\
   0.1 & 0.2 & 0.7
\end{array}
\right \}
\]
\end{center}
The results (fig.\ref{result1}) show that the channel capacity is
achieved after 20 iterations in the classical case, 7 iterations in
Matz's approach and 4 iterations in our case (with a precision of
$10^{-11}$).

\begin{figure}[!h]
\centerline{\epsfxsize=7cm\epsfysize=4cm\epsfbox{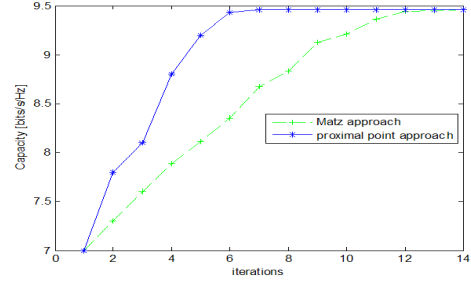}}\vspace*{-0.4cm}\caption{\footnotesize
\label{result2} Comparision between the 2 approaches in the case of
a Gaussian Bernouilli-Gaussian channel}
\end{figure}

A second example intends to characterize better the  efficiency of
our method in comparison with the one by Matz. In order to do so we
need a higher dimension problem. We have chosen the discretization
of  some continuous Gaussian Bernouilli-Gaussian channel  in order
to form a  transition channel matrix Q with higher dimensions. Such
a channel is defined as follows :
\begin{center}
$y_k=x_k+b_k+\gamma_k$
\end{center}
where
\begin{itemize}
\item
$b \sim \mathcal{N}(0,\sigma_b^2)$
\item
$\gamma_k=e_kg_k$ \hspace{0.5cm} with \hspace{1cm} e : Bernouilli(p)
sequence
\item
$g \sim \mathcal{N}(0,\sigma_g^2)$ \hspace{0.5cm} with \hspace{1cm}
$\sigma_b^2\ll \sigma_g^2$
\end{itemize}
Hence
\begin{center}
$y_k=x_k+n_k$
\end{center}
with \begin{center} $p(n_k)=(1-p) \mathcal{N}(0,\sigma_b^2)+p
\mathcal{N}(0,\sigma_b^2+\sigma_g^2)$
\end{center} The output $y_k$ has been discretized on $40$ values, and the input $x_k$ on $10$ values.
The results plotted on (fig.\ref{result2})  for parameters $(p=0.3,
\sigma_b=0.01, \sigma_g=1)$ show the acceleration of the
Blahut-Arimoto algorithm from 14 iterations in Matz's approach to 7
iterations in our method.
\section{Conclusions}
We have proposed geometrical interpretations and improvements  on
the Blahut-Arimoto (BA) algorithm for computing the capacity of
discrete memoryless channels (DMC). Based on the true proximal point
approach and solving the maximization problem with the conjugate
gradient method, we have accelerated the convergence rate of this
iterative algorithm compared to the aproach proposed by Matz which
is based on an approximation of the proximal point method.  We are
currently investigating the use of similar techniques for  improving
the convergence rate of other iterative algorithms.
\bibliographystyle{IEEEtran}
\bibliography{Icassp09ziad}
\end{document}